\documentclass[12pt]{article}
\usepackage{amsfonts}
\usepackage{amssymb}
\usepackage[mathcal]{euscript}
\usepackage[usenames,dvipsnames]{color}
\usepackage[rflt]{floatflt}
\usepackage{enumerate, subfigure}
\usepackage[charter]{mathdesign}
\usepackage{graphicx, amsmath,latexsym,times,color}
\usepackage{amsbsy}
\usepackage{epsfig}
\usepackage{verbatim}
\usepackage{gensymb}
\usepackage{geometry}
\usepackage{indentfirst}
\usepackage{float}
\usepackage[bf,singlelinecheck=off]{caption}
\usepackage{natbib}

\newtheorem{result}{Result}

\begin{document}

\thispagestyle{empty} \baselineskip 24pt

\begin{center}
{\LARGE \textbf{Test for Incremental Value of New Biomarkers Based on OR Rules}}
\end{center}

\begin{center}
Lu Wang\\[0pt]
Fred Hutchinson Cancer Research Center, Seattle, WA 98109, U.S.A. \\[0pt]
\textit{lwang235@fredhutch.org} \\[0pt]
~~\\[0pt]

Ying Huang\\[0pt]
Fred Hutchinson Cancer Research Center, Seattle, WA 98109, U.S.A. \\[0pt]
Department of Biostatistics, University of Washington, WA 98195, U.S.A.\\[0pt]~~
\textit{yhuang@fredhutch.org} \\[0pt]
~~\\[0pt]

Alexander R Luedtke\\[0pt]
Fred Hutchinson Cancer Research Center, Seattle, WA 98109, U.S.A. \\[0pt]
\textit{aluedtke@fredhutch.org} \\[0pt]
~~\\[0pt]

\end{center}

\baselineskip 24pt \clearpage
\thispagestyle{empty}

\begin{center}
{\large \textbf{Abstract}}
\end{center}
In early detection of disease, a single biomarker often has inadequate classification performance, making it important to identify new biomarkers to combine with the existing marker for improved performance. A biologically natural method to combine biomarkers is to use logic rules, e.g. the OR/AND rules. In our motivating example of early detection of pancreatic cancer, the established biomarker CA19-9 is only present in a subclass of cancer; it is of interest to identify new biomarkers present in the other subclasses and declare disease when either marker is positive. While there has been research on developing biomarker combinations using the OR/AND rules, the inference regarding the incremental value of the new marker within this framework is lacking and challenging due to a statistical non-regularity. In this paper, we aim to answer the inferential question of whether combining the new biomarker achieves better classification performance than using the existing biomarker alone, based on a nonparametrically estimated OR rule that maximizes the weighted average of sensitivity and specificity. We propose and compare various procedures for testing the incremental value of the new biomarker and constructing its confidence interval, using bootstrap, cross-validation, and a novel fuzzy p-value-based technique. We compare the performance of different methods via extensive simulation studies and apply them to the pancreatic cancer example. 

~~\newline

\clearpage\pagebreak\newpage \pagenumbering{arabic} \newlength{\gnat} %
\setlength{\gnat}{22pt} \baselineskip=\gnat

\baselineskip 24pt

\section{Introduction}
In early detection of disease, a single biomarker often has inadequate classification performance. Identifying new biomarkers to combine with established predictors (biomarkers)  for improved performance is an important research goal. For classification of binary diseases, examples of common modeling approaches for combining biomarkers include the likelihood-based logistic regression model, from which a marker combination score can be derived to subsequently generate a binary test based on a cutoff value. The use of logistic regression models has been well-studied in early detection; it yields optimal marker combination when the underlying risk model is correctly specified, but may otherwise have suboptimal classification performance. Another commonly used approach in the applied literature for combining markers in a binary test is the use of logic rules \citep{Etzioni2003}, e.g. the `OR/AND' rules \citep{Feng2010}, which consider combination rules to be the set of `or-and' combinations of threshold rules in each biomarker. To declare an individual as disease-positive, the OR rule requires that either one or the other marker passes its individual threshold, while the AND rule requires that both markers pass their thresholds.  For example, in early detection of pancreatic cancer, \cite{Tang2015} considered a two-marker panel that declares disease if either the established biomarker CA19-9 exceeds a threshold OR a new discovered glycan marker exceeds a threshold. In a prostate cancer screening study, \cite{Gann2002} showed that the addition of the ratio of free to total PSA (Prostate-Specific Antigen) within a specific total PSA range with the OR/AND rules could improve both specificity and sensitivity simultaneously relative to the conventional strategy based on total PSA alone.

Logic combination rules are desirable for combining biomarkers mostly because of their simplicity and interpretability. For example, the OR rule is often preferred due to its biological appeal in detecting cancer, which is typically heterogeneous and composed of different subclasses. If biomarkers from each subclass can be identified, an OR rule combining these biomarkers is expected to boost the overall sensitivity without sacrificing much specificity. On the other hand, the AND rule is considered to be useful when individual biomarkers for combination have very high sensitivity and low specificity. Our research in this paper is motivated by the development of biomarker combinations to improve early detection of pancreatic cancer. The current best marker for early detection of pancreatic cancer is the CA19-9 test, which detects the sialyl-Lewis A (sLeA) glycan. sLeA levels are not elevated in 25\% of pancreatic cancers due to factors such as genetic inability. It is of interest to discover glycans other than sLeA that are overproduced in some pancreatic cancers that are low in sLeA. It is hoped that these glycans, when combined with CA19-9 using the OR rule (i.e. declaring a case if levels of either CA19-9 or the new marker are elevated), can improve the classification performance over CA19-9 alone \citep{Tang2015}. In other words, an important question that needs to be addressed here is whether the new biomarker has significant incremental value when combined with an established biomarker, compared to using the established biomarker alone. While several authors have conducted statistical research in combination of biomarkers using the OR/AND rules, the focus in the past has been mainly on algorithm development for finding the best combination instead of on making inference about the new biomarker's incremental value. For example, \cite{Baker2000} proposed a nonparametric multivariate algorithm that extended the idea of receiver operating characteristic (ROC) cutpoints to multivariate positivity regions in order to find the optimal ROC curve. \cite{Etzioni2003} considered classifying prostate cancers using OR/AND rules that combined total PSA with the ratio of free/total PSA; LOGIC regression \cite{Ruczinski2003} was performed to find the best logic rule that maximizes the cross-validated weighted sum of sensitivity and specificity. Statistical research to answer the inferential question about the incremental value of a new biomarker, however, is lacking. As we will show next, this is a challenging problem due to the non-regularity of the incremental estimator under the null hypothesis (i.e. when the new biomarker has no incremental value over the established biomarker). In this paper, we aim to fill this gap. We will propose and compare various strategies for making inference regarding a new biomarker's incremental value over an established biomarker. We consider a simple OR rule in this paper for combining the established marker with the new marker, motivated by the pancreatic cancer example. However, the technique can be generalized to AND rule or OR/AND rule combinations. 

This paper is organized as follows: In Section 2, we present an estimator for an OR rule that maximizes the weighted average of sensitivity and specificity, based on which the incremental value of the new marker is estimated. We propose different procedures for testing the significance of the incremental value and construct its confidence interval, utilizing the bootstrap, cross-validation, and a novel fuzzy p-value technique. In Section 3, we conduct extensive simulation studies to compare performance of different methods. The application of the methods developed to the pancreatic cancer example is illustrated in Section 4. We make concluding remarks in Section 5.

\section{Methodology}
\label{Method}

Let $D$ be a binary disease outcome, with value $1$ for diseased and $0$ for non-diseased. Let $X$ be an established biomarker (predictor) for predicting $D$ and let $Y$ be a new biomarker that we are interested in evaluating. The objective is to test whether combination of $Y$ with $X$ based on the OR rule offers any incremental value in classification performance over $X$ alone and to estimate the incremental value. Suppose a case is declared if either $X$ or $Y$ is elevated, i.e., $X>c_{1}$ or $Y>c_{2}$, for some thresholds $c_{1}$ and $c_{2}$. We define sensitivity as $Sen\left( c_{1},c_{2}\right) =P\left(X>c_{1}\text{ }OR\text{ }Y>c_{2}\mid D=1\right) $ and specificity as $Spe\left( c_{1},c_{2}\right) =P\left( X\leq c_{1}\text{ }AND\text{ }Y\leq c_{2}\mid D=0\right) $. For a test based on the biomarker $X$ alone, sensitivity and specificity are defined as $Sen\left( c\right) =P\left( X>c\mid D=1\right) $ and $Spe\left( c\right) =P\left( X\leq c\mid D=0\right)$, respectively, for some threshold $c$. To characterize the incremental performance of the new biomarker $Y$, we consider the weighted average of sensitivity and specificity as an overall summary measure of a model's performance \citep{Han2011}, for pre-specified weight $w\in (0,1)$, for either the model based on $X$ alone or the model based on the combination of $X$ and $Y$. The most common special case of the weighted average of sensitivity and specificity is the Youden's index \citep{Youden1950}, which weights a model's sensitivity and specificity equally. This index will be adopted in our numerical studies in this paper. In practice, weights for sensitivity/specificity can be chosen to reflect the relative importance of not missing the detection of a case versus not making false positive detection of a control, from a cost/benefit perspective. We define the incremental value of the new marker $Y$ as the increase in the maximum value of the weighted average of sensitivity and specificity using the OR rule combining $X$ and $Y$ compared with the rule using $X$ alone, i.e.
\begin{equation}
\theta =\underset{c_{1},c_{2}}{\max }\left[ \omega Sen\left(
c_{1},c_{2}\right) +\left( 1-\omega \right) Spe\left( c_{1},c_{2}\right) %
\right] -\underset{c}{\max }\left[ \omega Sen\left( c\right) +\left(
1-\omega \right) Spe\left( c\right) \right]  \label{eqn1},
\end{equation} 
with some weight $\omega \in \left( 0,1\right)$. The incremental value is always nonnegative.

We note a connection between the maximizer of the weighted average of sensitivity and specificity and a logic regression risk model \citep{Ruczinski2003}. Specifically, suppose the risk of the disease conditional on the established biomarker $X$ follows a logic regression model
\begin{equation}
P\left(D=1 \mid X\right)=\alpha_{0} + \alpha_{1}I\left(X>c\right). \label{logic_model_x}
\end{equation}
When conditions in {\bf Result 1} below are satisfied, the threshold $c$ in (\ref{logic_model_x}) will be the one that maximizes the weighted average of sensitivity and specificity in a binary classification rule based on $X$ alone. Similarly, suppose the risk of disease conditional on $X$ and $Y$ follows a logic regression model   
\begin{equation}
P\left(D=1 \mid X,Y\right)=\beta_{0} + \beta_{1}I\left(X>c_{1}\text{ }OR\text{ }Y>c_{2}\right) \label{logic_model}.
\end{equation}
When conditions in {\bf Result 2} below are satisfied, the thresholds $c_1$ and $c_2$ in (\ref{logic_model}) will be the corresponding thresholds for $X$ and $Y$ that maximize the weighted average of sensitivity and specificity in an OR rule.

\begin{result}
	For a binary rule based on $X$ that classifies an observation as diseased if $X>\delta$, suppose the threshold value $c^{\ast}$ is the unique maximizer of the weighted average of sensitivity and specificity, i.e., 
	$c^{\ast}=\arg \underset{\delta}{\max }\left[\omega Sen\left(\delta\right)+\left(1-\omega \right) Spe\left(\delta\right)\right]$. If (\ref{logic_model_x}) holds, the CDF of $X$ is not flat in a neighborhood of $X$, $\omega \frac{\alpha_{0}+\alpha_{1}}{P\left(D=1\right)}-\left(1-\omega\right) \frac{1-\alpha_{0}-\alpha_{1}}{P\left(D=0\right)}> 0$ and $\omega \frac{-\alpha_{0}}{P\left(D=1\right)}+\left(1-\omega\right) \frac{1-\alpha_{0}}{P\left(D=0\right)}> 0$,  then $c^*$ equals the parameter $c$ indexing \eqref{logic_model_x}.
\end{result}
{\bf Proof:}

Assume risk model (\ref{logic_model_x}) holds. We can show that for $\delta\ge c$,
\begin{eqnarray*}
	Sen(\delta)=\frac{P(X>\delta, D=1)}{P(D=1)}=\frac{\int_{\delta^+}^{\infty}P(D=1|x)dF_X(x)}{P(D=1)}=\frac{(\alpha_0+\alpha_1)(1-F_X(\delta))}{P(D=1)}, 
\end{eqnarray*}
and that for $\delta\le c$,
\begin{eqnarray*}
	Sen(\delta)=\frac{P(D=1)-P(X\le \delta, D=1)}{P(D=1)}=1-\frac{\int_{-\infty}^{\delta}P(D=1|x)dF_X(x)}{P(D=1)}=1-\frac{\alpha_0(1-F_X(\delta))}{P(D=1)}. 
\end{eqnarray*}
Similarly, $Spe(\delta)$ equals $1-\frac{(1-\alpha_0-\alpha_1)(1-F_X(\delta))}{P(D=0)}$ for $\delta\ge c$ and equals $\frac{(1-\alpha_0)(1-F_X(\delta))}{P(D=0)}$ for $\delta\le c$.

Let $A\left(\delta\right)=\omega Sen\left(\delta\right)+\left(1-\omega \right) Spe\left(\delta\right)$,
then for $\delta>c$,
\begin{equation*}
A\left(\delta\right)-A\left(c\right)=\left[\omega \frac{\alpha_{0}+\alpha_{1}}{P\left(D=1\right)}-\left(1-\omega\right) \frac{1-\alpha_{0}-\alpha_{1}}{P\left(D=0\right)} \right]\left[F_{X}\left(c\right)-F_{X}\left(\delta\right)\right];
\end{equation*}
and for $\delta<c$,
\begin{equation*}
A\left(\delta\right)-A\left(c\right)=\left[\omega \frac{-\alpha_{0}}{P\left(D=1\right)}+\left(1-\omega\right) \frac{1-\alpha_{0}}{P\left(D=0\right)}\right]\left[F_{X}\left(\delta\right)-F_{X}\left(c\right)\right],
\end{equation*}
where $F_{X}$ is the CDF of the biomarker $X$. Therefore, $c^{\ast}=c$ if $\omega \frac{\alpha_{0}+\alpha_{1}}{P\left(D=1\right)}-\left(1-\omega\right) \frac{1-\alpha_{0}-\alpha_{1}}{P\left(D=0\right)}> 0$ and $\omega \frac{-\alpha_{0}}{P\left(D=1\right)}+\left(1-\omega\right) \frac{1-\alpha_{0}}{P\left(D=0\right)}> 0$.

\begin{result}
	For an OR rule based on $X$ and $Y$ that classifies an observation as diseased if $X>\delta_1$ or $Y>\delta_2$, suppose the threshold value ($c_1^{\ast}$, $c_2^{\ast}$) for $X$ and $Y$ is the unique maximizer of the weighted average of sensitivity and specificity, i.e., $\left(c_{1}^{\ast},c_{2}^{\ast}\right)=\arg \underset{\delta_{1},\delta_{2}}{\max }\left[\omega Sen\left(\delta_{1},\delta_{2}\right)+\right.$ \\ $\left.\left(1-\omega \right) Spe\left(\delta_{1},\delta_{2}\right)\right]$. If condition (\ref{logic_model}) holds, the CDF of $(X,Y)$ is not flat in the neighborhood of $c_1, c_2$, $\omega \frac{\beta_{0}+\beta_{1}}{P\left(D=1\right)}-\left(1-\omega\right) \frac{1-\beta_{0}-\beta_{1}}{P\left(D=0\right)}> 0$ and $\omega \frac{-\beta_{0}}{P\left(D=1\right)}+\left(1-\omega\right) \frac{1-\beta_{0}}{P\left(D=0\right)}>0$, then $(c_1^{\ast}, c_2^{\ast})$ equals parameters $(c_1,c_2)$ indexing (\ref{logic_model}).
\end{result}
{\bf Proof:}

Assume risk model (\ref{logic_model}) holds, we can derive $Sen(\delta_1,\delta_2)$ and $Spe(\delta_1,\delta_2)$ for various combinations of $\delta_1$ and $\delta_2$. For example, for $\delta_1\le c_1$ and $\delta_2\le c_2$, 
\begin{eqnarray*}
	&&Sen(\delta_1,\delta_2)=1-\frac{P(X\le \delta_1 \;\&\; Y\le \delta_2, D=1)}{P(D=1)} \cr
	&=&1-\frac{\int_{-\infty}^{\delta_1}\int_{-\infty}^{\delta_2}\beta_0dF_{X,Y}(x,y)}{P(D=1)}
	=1-\frac{\beta_0F_{X,Y}(\delta_1,\delta_2)}{P(D=1)},
\end{eqnarray*}	
\begin{eqnarray*}
	&&Spe(\delta_1,\delta_2)=\frac{P(X\le \delta_1 \;\&\; Y\le \delta_2, D=0)}{P(D=0)} \cr
	&=&\frac{\int_{-\infty}^{\delta_1}\int_{-\infty}^{\delta_2}\beta_0dF_{X,Y}(x,y)}{P(D=0)}
	=\frac{\beta_0F_{X,Y}(\delta_1,\delta_2)}{P(D=0)}.
\end{eqnarray*}	
Let $B\left(\delta_{1},\delta_{2}\right)=\omega Sen\left(\delta_{1},\delta_{2}\right)+\left(1-\omega \right) Spe\left(\delta_{1},\delta_{2}\right)$. We can show that \\for $\delta_{1}\leq c_{1}$, $\delta_{2}< c_{2}$,
\begin{equation*}
B\left(\delta_{1},\delta_{2}\right)-B\left(c_{1},c_{2}\right)=\left[\omega \frac{-\beta_{0}}{P\left(D=1\right)}+\left(1-\omega\right) \frac{1-\beta_{0}}{P\left(D=0\right)} \right]\left[F_{X,Y}\left(\delta_{1},\delta_{2}\right)-F_{X,Y}\left(c_{1},c_{2}\right)\right];
\end{equation*}
for $\delta_{1}> c_{1}$, $\delta_{2}< c_{2}$,
\begin{eqnarray*}
	B\left(\delta_{1},\delta_{2}\right)-B\left(c_{1},c_{2}\right)=\left[\omega \frac{-\beta_{0}}{P\left(D=1\right)}+\left(1-\omega\right) \frac{1-\beta_{0}}{P\left(D=0\right)} \right]\left[F_{X,Y}\left(c_{1},\delta_{2}\right)-F_{X,Y}\left(c_{1},c_{2}\right)\right] \\
	+\left[\omega \frac{\beta_{0}+\beta_{1}}{P\left(D=1\right)}-\left(1-\omega\right) \frac{1-\beta_{0}-\beta_{1}}{P\left(D=0\right)} \right]\left[F_{X,Y}\left(c_{1},\delta_{2}\right)-F_{X,Y}\left(\delta_{1},\delta_{2}\right)\right] ;
\end{eqnarray*}
for $\delta_{1} \leq c_{1}$, $\delta_{2} > c_{2}$,
\begin{eqnarray*}
	B\left(\delta_{1},\delta_{2}\right)-B\left(c_{1},c_{2}\right)=\left[\omega \frac{-\beta_{0}}{P\left(D=1\right)}+\left(1-\omega\right) \frac{1-\beta_{0}}{P\left(D=0\right)} \right]\left[F_{X,Y}\left(\delta_{1},c_{2}\right)-F_{X,Y}\left(c_{1},c_{2}\right)\right] \\
	+\left[\omega \frac{\beta_{0}+\beta_{1}}{P\left(D=1\right)}-\left(1-\omega\right) \frac{1-\beta_{0}-\beta_{1}}{P\left(D=0\right)} \right]\left[F_{X,Y}\left(\delta_{1},c_{2}\right)-F_{X,Y}\left(\delta_{1},\delta_{2}\right)\right] ;
\end{eqnarray*}
and for $\delta_{1} > c_{1}$, $\delta_{2} > c_{2}$,
\begin{equation*}
B\left(\delta_{1},\delta_{2}\right)-B\left(c_{1},c_{2}\right)= \left[\omega \frac{\beta_{0}+\beta_{1}}{P\left(D=1\right)}-\left(1-\omega\right) \frac{1-\beta_{0}-\beta_{1}}{P\left(D=0\right)} \right]\left[F_{X,Y}\left(c_{1},c_{2}\right)-F_{X,Y}\left(\delta_1,\delta_2\right)\right], 
\end{equation*}
where $F_{X,Y}$ is joint CDF of biomarkers $X$ and $Y$. Therefore, $c_{1}^{\ast}=c_{1}$, $c_{2}^{\ast}=c_{2}$ if $\omega \frac{\beta_{0}+\beta_{1}}{P\left(D=1\right)}-\left(1-\omega\right) \frac{1-\beta_{0}-\beta_{1}}{P\left(D=0\right)}> 0$ and $\omega \frac{-\beta_{0}}{P\left(D=1\right)}+\left(1-\omega\right) \frac{1-\beta_{0}}{P\left(D=0\right)}> 0$.

In general, even when the actual disease risk model conditional on biomarker(s) may not follow the conditions specified in Results 1 and 2, it is still appealing to identify classification rules based on $X$ alone or an OR combination of $X$ and $Y$ by maximizing the weighted average of sensitivity and specificity, given that the weighted average of sensitivity and specificity is a clinically meaningful operational criterion of practical interest. So is the estimation of the incremental value of $Y$ based on difference in model performance, as defined in equation (\ref{eqn1}).

\subsection{Inference}

\subsubsection{Estimation}
To estimate the incremental value $\theta$, we consider nonparametric estimators of the classification rule based on either $X$ alone or combinations of $X$ and $Y$. In particular, we estimate threshold(s) in the corresponding rules by maximizing the weighted average of nonparametric estimates of sensitivity and specificity. Let subscripts $D$ and $\bar{D}$ indicate case and control status, respectively, such that $X_D$ and $Y_D$ indicate biomarker measurements among cases and $X_{\bar{D}}$ and $Y_{\bar{D}}$ indicate biomarker measurements among controls. Let $n_D$ and $n_{\bar{D}}$ be sample sizes for cases and controls, respectively. We compute $\hat{c}$ as the maximizer in $c$ of $w\sum_{i=1}^{n_D}I(X_{Di}>c)/{n_D}+(1-w)\sum_{j=1}^{n_{\bar{D}}}I(X_{\bar{D}j}\le c)/n_{\bar{D}}$. Similarly we compute $\hat{c}_1$ and $\hat{c}_2$ as the maximizers in $c_1$ and $c_2$ of $w\sum_{i=1}^{n_D}I(X_{Di}>c_1 \text{ }OR\text{ } Y_{Di}>c_2)/{n_D}+(1-w)\sum_{j=1}^{n_{\bar{D}}}I(X_{\bar{D}j}\le c_1 \text{ }AND\text{ } Y_{\bar{D}j}\le c_2)/n_{\bar{D}}$. Based on $\hat{c}$, $\hat{c}_1$ and $\hat{c}_2$, we then estimate $\theta$ nonparametrically as
\begin{align*}
\hat{\theta}=\,&\left[\frac{w}{n_D}\sum_{i=1}^{n_D}I(X_{Di}>\hat{c}_1 \text{ }OR\text{ } Y_{Di}>\hat{c}_2)+\frac{1-w}{n_{\bar{D}}}\sum_{j=1}^{n_{\bar{D}}}I(X_{\bar{D}j}\le \hat{c}_1 \text{ }AND\text{ } Y_{\bar{D}j}\le \hat{c}_2)\right]\\
&-\left[\frac{w}{n_D}\sum_{i=1}^{n_D}I(X_{Di}>\hat{c})+\frac{1-w}{n_{\bar{D}}}\sum_{j=1}^{n_{\bar{D}}}I(X_{\bar{D}j}\le \hat{c})\right].
\end{align*} 
Note however this `na\"{i}ve' estimator estimates the rule and its performance from the same dataset and thus is subject to overfitting bias. To reduce overfitting, a K-fold cross-validation method can be adopted instead. In performing cross-validation, first the dataset is split into $K$ mutually exclusive and exhaustive subsets stratified on case/control status. Each time, one of the $K$ subsets is used as the test set and the remaining $K-1$ subsets are combined together to form a training set. The thresholds are estimated based on the training set and then they are used to obtain the incremental value estimator based on the $k^{th}$ test set, denoted by $\hat{\theta}_{k}$. The cross-validated estimator of incremental value is produced by taking average of the resulting $K$ estimators, i.e., 
\begin{equation*}
\hat{\theta}_{CV}=\frac{1}{K}\sum_{k=1}^{K}\hat{\theta}_{k}.
\end{equation*}

Next we investigate approaches to test the hypothesis that $Y$ has significant incremental value when combined with $X$ through an OR rule, i.e., to test 
\begin{equation*}
H_{0}:\theta =0\text{ versus }H_{1}:\theta >0,
\end{equation*}
as well as approaches to construct confidence interval of the incremental value. We will investigate bootstrap-based approaches for inference as well as proposing a novel fuzzy P-value-based approach.

\subsubsection{Bootstrap Approach}

\begin{figure}[H] 	
	\caption{Distributions of $\hat{\theta}$ and $\hat{\theta}_{CV}$ based on 1000 simulated datasets, $\hat{\theta}^{\ast}$ and $\hat{\theta}^{\ast}_{CV}$ over 500 bootstrap samples, for $\theta=0$, 0.01 and 0.05.}
	\label{Figure}
	\centering \scalebox{0.6}  { 
		\includegraphics{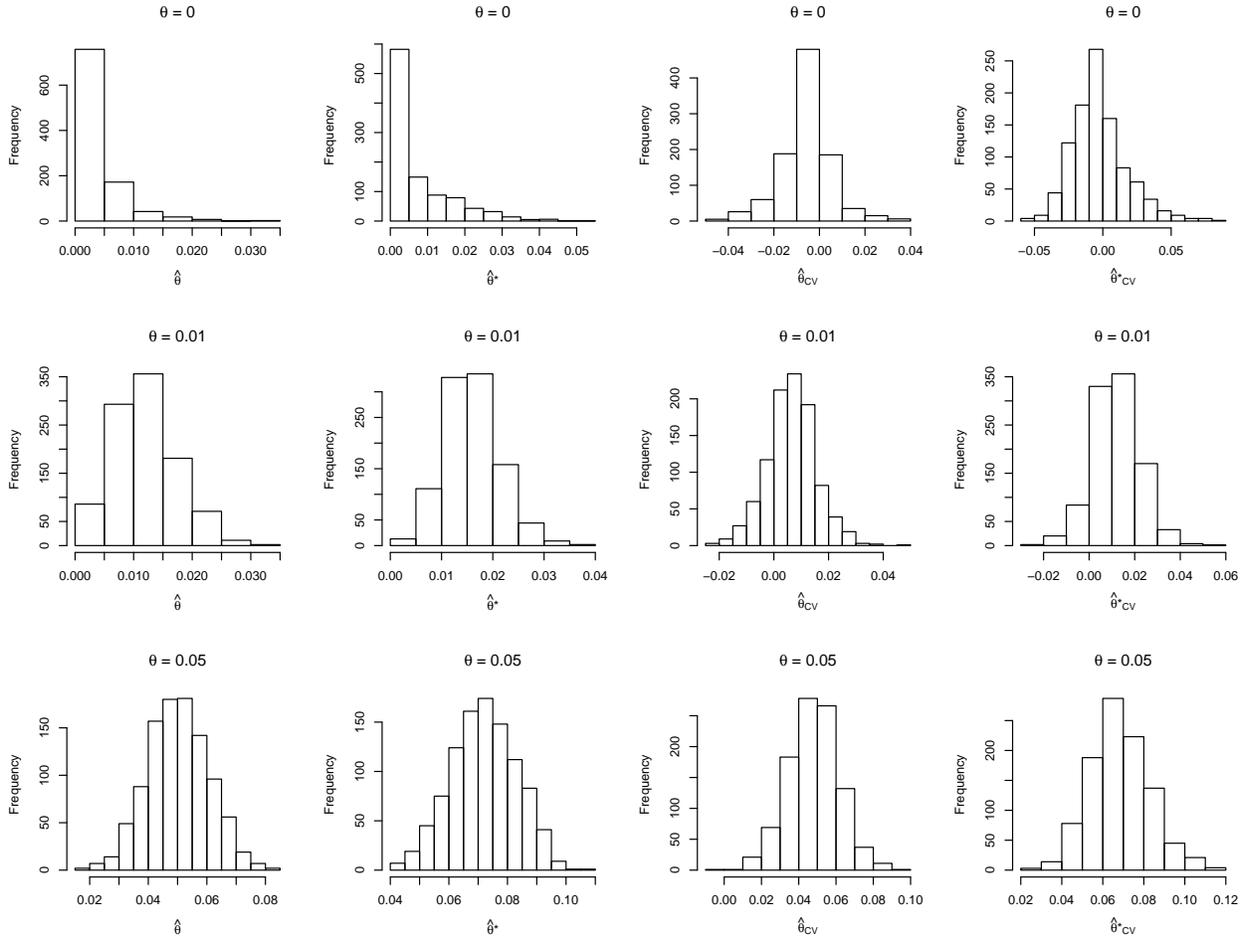}}
\end{figure}

A challenge with the test of incremental value in this problem setting is the non-regularity of the incremental value estimator under the null hypothesis. In other words, the na\"{i}ve nonparametric estimator $\hat{\theta}$ is not asymptotically normal so the standard testing procedure based on asymptotic normality of the test statistics is not applicable here. Figure \ref{Figure} presents numerical examples of distribution of $\hat{\theta}$ for various $\theta$ values. When the null hypothesis is true ($\theta=0$), $\hat{\theta}$ is heavily right-skewed with a peak at zero. The distribution of $\hat{\theta}$ gradually approaches normality as $\theta$ moves away from zero. 

A commonly used approach for performing hypothesis tests is to construct bootstrap \citep{Efron1994} confidence intervals (CI) for an estimand and evaluate whether the CI covers the parameter value specified in the null hypothesis. Here we investigate different bootstrap methods to perform the hypothesis test about the incremental value of $Y$. Both empirical and percentile bootstrap methods are considered. Suppose we have a data set of size $n$, from which we draw $B$ random samples of size $n$ with replacement, stratified on case/control status. Let $\hat{\theta}$ and $\hat{\theta}^{\ast }$ be the nonparametric incremental value estimates based on the original data set and the bootstrap samples, respectively. The one-sided $\left(1-\alpha \right)\times100\%$ empirical bootstrap confidence intervals are constructed as $\left[ 2\hat{\theta}-\hat{\theta}_{1-\alpha}^{\ast },\infty \right)$, where  $\hat{\theta}_{1-\alpha}^{\ast }$ denotes the $1-\alpha$ quantile of $\hat{\theta}^{\ast }$. The one-sided $\left(1-\alpha \right)\times100\%$ percentile bootstrap confidence intervals are constructed as $\left[\hat{\theta}_{\alpha}^{\ast },\infty \right)$. The one-sided test for the incremental value being greater than zero can be based on whether the lower bound of the $(1-\alpha)\times 100\%$ one-sided bootstrap CI is above zero. Percentile bootstrap CI has been widely used in biomarker research for characterizing and comparing biomarker performances. However, its validity requires symmetry in the distribution of the estimator \citep{Van1998}, which is clearly violated under the null hypothesis in our problem setting based on $\hat{\theta}$. In contrast, the rationale behind the empirical bootstrap is to approximate the distribution of  $\hat{\theta} - \theta$ by the distribution of $\hat{\theta}^{\ast}-\hat{\theta}$. From Figure 1, the right tails of the distributions for $\hat{\theta}$ and $\hat{\theta}^{\ast}$ agree reasonably well, suggesting the potential of testing the incremental value based on the lower confidence limit of the one-sided empirical bootstrap CI based on $\hat{\theta}$. Nonetheless, we emphasize that we do not currently have theory supporting the validity of the bootstrap under the null, and therefore our simulation will serve as preliminary evidence for or against its validity. Hereafter we refer to the approaches based on empirical bootstrap CI or percentile bootstrap CI of $\hat{\theta}$ as EB and PB, respectively.  

In addition to bootstrap CIs based on the na\"{i}ve estimate $\hat{\theta}$, we also consider the construction of Wald-type CIs based on $\hat{\theta}_{CV}$, the cross-validated incremental estimate. From Figure \ref{Figure}, the distribution of the cross-validated estimate $\hat{\theta}_{CV}$ is approximately normal under both the null and the alternative hypotheses, suggesting the validity of using Wald method for CI construction, where the cross-validated estimate is computed on each bootstrap sample. The one-sided and two-sided $\left(1-\alpha \right)\times100\%$ Wald CIs are $\left[ \hat{\theta}_{CV}-z_{1-\alpha }SD\left(\hat{\theta}_{CV}^{\ast }\right),\infty \right)$ and $\left[ \hat{\theta}_{CV}-z_{1-\alpha/2 }SD\left(\hat{\theta}_{CV}^{\ast }\right), \hat{\theta}_{CV}+z_{1-\alpha/2 }SD\left(\hat{\theta}_{CV}^{\ast }\right)\right]$, respectively, where $z_{1-\alpha}$ denotes the $1-\alpha$ quantile of a standard normal random variable, $SD\left(\hat{\theta}_{CV}^{\ast }\right)$ is the standard deviation of bootstrap cross-validated estimate $\hat{\theta}_{CV}^{\ast }$. Hereafter we refer to the approach based on Wald CI of $\hat{\theta}$ as Wald.CV. 

\subsubsection{Fuzzy P-value Approach}
In this section, we propose a novel test for incremental value of $Y$ that leverages the fact that $n^{1/2}\hat{\theta}$ will converge to zero under the null given some regularity conditions. Let $\psi=\underset{c_{1},c_{2}}{\max }\left[ \omega Sen\left(c_{1},c_{2}\right) +\left( 1-\omega \right) Spe\left( c_{1},c_{2}\right) \right]$ and $\phi=\underset{c}{\max }\left[ \omega Sen\left( c\right) +\left(1-\omega \right) Spe\left( c\right) \right]$. Under some regularity conditions, the key of which is that the $c$ maximizing $\omega Sen\left( c\right) +\left(1-\omega \right) Spe\left( c\right)$ is unique, one can show that the NPMLE $\hat{\phi}$ satisfies the following asymptotically linear expansion
\begin{align}
\hat{\phi}-\phi=\,& \frac{1}{n}\sum_{i=1}^n \Bigg(\omega\frac{I\{D_i=1\}}{P(D=1)}\left[I\{X_i>c^{\ast}\}-P(X>c^{\ast}|D=1)\right] \nonumber \\
&+(1-\omega)\frac{I\{D_i=0\}}{P(D=0)}\left[I\{X_i\le c^{\ast}\}-P(X\le c^{\ast}|D=0)\right]\Bigg) + o_P(n^{-1/2}), \label{eq:phiexp}
\end{align}
where above $o_P(n^{-1/2})$ represents a term that converges to zero in probability once multiplied by $n^{1/2}$. The term in the sum represents the canonical gradient of the parameter $\phi$ \citep{Bickel1993}. Under the i.i.d. assumption, $\sqrt{n}\left(\hat{\phi}-\phi \right) \rightsquigarrow N\left(0,\sigma_{\phi}^{2}\right)$ as the sample size goes to infinity, where $\sigma_{\phi}^{2}$ is the efficiency bound for regular and asymptotically linear estimators for $\phi$ within the nonparametric model. Here we have made the simplifying assumption that the data are a sample of $n$ independent and identically distributed (i.i.d.) observations, so that standard efficiency theory can be applied. Nonetheless, if a fixed number of cases and controls are sampled, then the dominant term above breaks into the sum of an empirical mean over cases and an empirical mean over controls, and the remainder term will remain negligible. Hence, central limit theorem results can be obtained in that case as well. Under similar regularity conditions, the key of which is that the $c_1,c_2$ in the closure of the support of $X,Y$ maximizing $\omega Sen\left(c_{1},c_{2}\right) +\left( 1-\omega \right) Spe\left( c_{1},c_{2}\right)$ are unique,
\begin{align}
\hat{\psi}-\psi=\,& \frac{1}{n}\sum_{i=1}^n \Bigg(\omega\frac{I\{D_i=1\}}{P(D=1)}\left[I\{X_i>c_1^{\ast}\;OR\;Y_i>c_2\}-P(X>c_1^{\ast}\;OR\;Y>c_2^{\ast}|D=1)\right] \nonumber \\ \notag
&+(1-\omega)\frac{I\{D_i=0\}}{P(D=0)}\left[I\{X_i\le c_1^{\ast}\;AND\;Y_i\le c_2\}-P(X\le c_1^{\ast}\;AND\;Y\le c_2^{\ast}|D=0)\right]\Bigg) \\ 
& + o_P(n^{-1/2}).  \label{eq:psiexp}
\end{align}
Hence, $\sqrt{n}\left(\hat{\psi}-\psi \right) \rightsquigarrow N\left(0,\sigma_{\psi}^{2}\right)$. Under the assumptions needed for \eqref{eq:phiexp} and \eqref{eq:psiexp} to hold, the null hypothesis that $\theta=0$ implies that $c_2^*$ falls at the upper edge of the support for $Y$, in the sense that it is equal to the smallest number $c_2$ such that $P(Y>c_2)=0$. In this case, the right-hand sides of \eqref{eq:phiexp} and \eqref{eq:psiexp} are equivalent up to an $o_P(n^{-1/2})$ term, so that
\begin{align*}
\textit{Under the null:}&\hspace{2em}n^{1/2}\hat{\theta} = n^{1/2}[\hat{\theta} - \theta] = n^{1/2}[\hat{\psi} - \psi] - n^{1/2}[\hat{\phi}-\phi] = o_P(1).
\end{align*}
Now, using that $\hat{\psi}$ and $\hat{\phi}$ are consistent estimators, we also have that
\begin{align*}
\textit{Under a fixed alternative:}&\hspace{2em}n^{1/2}\hat{\theta}\textnormal{ diverges in probability,}
\end{align*}
i.e., for any fixed $t>0$, $P(n^{1/2}\hat{\theta}>t)\rightarrow 1$ as $n\rightarrow\infty$.

We now use these facts to introduce a fuzzy p-value. Let $F$ denote a cumulative distribution function for a continuous random variable on $\mathbb{R}$. By the above two facts, if $Z$ is a random variable with cumulative distribution function $F$, under the null $p(Z)\equiv F(Z-n^{1/2}\hat{\theta})$ converges to a standard uniform random variable, whereas, under a fixed alternative, $p(Z)$ converges in probability to $1$. Note that $p(Z)$ is a valid fuzzy p-value according to the definition given in \cite{Geyer2005}. To generate a concrete decision, one could sample $Z$ from $F$ and reject if $p(Z)> \alpha$. Under the null hypothesis, the null will reject with probability approaching $\alpha$. In our simulation, we use $\hat{\theta}_{CV}$ in computation of the fuzzy p-value to minimize over-fitting bias in small sample size; we let $F$ be equal to the CDF of the normal distribution $N\left(0,\hat{\sigma}_{\psi}^{2}\right)$, where the estimate $\hat{\sigma}_{\psi}^{2}$ of the variance $\sigma_{\psi}^2$ is estimated via the bootstrap.

\section{Simulation Study}
\label{Simulation}

In this section, we conduct simulation studies to compare the performance of the methods described in Section \ref{Method} for testing and making inference about a new biomarker's incremental value. Here we consider equally weighted sensitivity and specificity as the classification performance measure and define $\phi$ and $\psi$ as the optimal average sensitivity and specificity based on an established marker alone or based on its combination with a new marker with an OR rule. The incremental value $\theta$ is then defined as the difference between $\psi$ and $\phi$. 

Let $D$ be a binary disease outcome. Let $X$ and $Y$ be two biomarkers that are independently distributed; each
follows the standard normal distribution. We consider two scenarios where the underlying true risk model is (1) a logic model for the risk of $D$ conditional on $X$ and $Y$: $P\left(D=1\mid X,Y\right) =\beta _{0}+\beta _{1}I\left( X>c_{1}\text{ }OR\text{ }Y>c_{2}\right)$ with thresholds $c_{1}$, $c_{2}$ and parameters $\beta_{0}$, $\beta _{1}$. , and (2) a logistic risk model $logit\left(P\left(D=1\mid X,Y\right)\right) = log(1/9) + \beta_{1}^{\ast}X+\beta_{2}^{\ast}Y$, yet investigators have adopted the simple OR rule to combine $X$ and $Y$. In both scenarios, we set disease prevalence $P\left(D=1\right)=0.1$ with appropriate selection of $c_{1}$, $c_{2}$, $\beta_{0}$, $\beta_{1}$, $\beta_{1}^{\ast}$ and $\beta _{2}^{\ast}$. For the logic model, the true incremental value of $Y$ is $0$ when $c_{2}=\infty$, otherwise it is greater than $0$. We choose two different values of the threshold for biomarker $X$, $c_{1}=0$ and $0.842$, which correspond to the $50^{th}$ and $80^{th}$ percentiles of the standard normal distribution, respectively. In the following, we select different values of $\beta _{0}$, $\beta _{1}$ or $\beta_{1}^{\ast}$, $\beta _{2}^{\ast}$ to achieve a wide variety of classification performance based on $X$ alone ($\phi$) or based on the OR combination of $X$ and $Y$ ($\psi$). Here we consider the scenario where classification performance of the established biomarker $X$ is not very weak ($\phi \geq 0.6$), which is likely to be true in practice. A range of incremental values \{$\theta=0$, $0.05$, or $0.1$\} are considered in our simulation studies.  

We consider case/control samples with equal numbers of cases and controls $n_{D} = n_{\bar{D}}=100$, $250$ or $500$ randomly sampled from the population, based on which we compute the na\"{i}ve performance estimate $\hat{\theta}$ and the cross-validated estimate $\hat{\theta}_{CV}$ based on 10-fold cross-validation. For each setting, we evaluate bias of $\hat{\theta}$ and $\hat{\theta}_{CV}$. We compare four different methods to perform the one-sided test for incremental value with respect to type I error rate and power: i) EB: the method based on empirical bootstrap CI using $\hat{\theta}$, ii) PB: the method based on percentile bootstrap CI using $\hat{\theta}$, iii) Wald.CV, Wald test using $\hat{\theta}_{CV}$, and iv) the fuzzy p-value approach. In addition, we examine coverage of $95\%$ two-sided CI of incremental value using the Wald.CV method. In each setting, 1000 Monte-Carlo simulations are conducted with 1000 bootstrap replicates constructed stratified on case/control status. The simulation results for various scenarios and sampling size at each fixed $\theta$ are summarized in Tables \ref{Est}, \ref{RejectRate}, and \ref{Coverage}. Corresponding results under local alternative where $\theta\propto n_D^{-1/2}$ are presented in the appendix (Table 4).  

From Tables \ref{Est} and Table 4 (appendix), under both correctly specified and mis-specified underlying models, we see that the na\"{i}ve estimator tends to overestimate the new biomarker's performance when the true incremental value is small (e.g. $\theta \le 0.05$); the overestimation issue becomes less severe as sample size increases as well as for settings with better performance of $X$ alone (i.e. settings with larger $\phi$ value). Using cross-validation in general corrects this overestimation problem and can lead to small attenuation in some settings. 

From Table \ref{RejectRate}, when the null hypothesis is true, the test based on PB often has inflated type I error, whereas the corresponding test based on EB typically has type I error rate close to or smaller than the nominal level (e.g. when biomarker $X$ itself has good performance). Wald.CV based on cross-validated $\theta$ estimate controls type I error, while it is often more conservative than EB. The fuzzy p-value method works very well with type I error fairly close to the nominal level for all settings. When the alternative hypothesis is true, the test based on EB in general has better or comparable power compared to other tests; the performance of Wald.CV and the fuzzy p-value method are more or less comparable to each other and their relative performance varies across settings.  

From Tables \ref{Coverage} and Table 4 (appendix), for the purpose of constructing two-sided CI for the incremental value, the Wald.CV approach is satisfactory for both underlying models. It clearly shows that the Wald.CV two-sided CIs have coverage either close to or slightly larger than the nominal level. Under the mis-specified model and alternative hypothesis, although the two-sided CIs can have slight undercoverage for the smaller sample size, its coverage approaches the nominal level as when the sample size is large enough, i.e., $n_{D}=n_{\bar{D}}=500$. 

Overall, we observe a similar pattern on performance comparison among different approaches under the (correctly specified) logic and (mis-specified) logistic risk models. For testing the significant incremental value of a new biomarker, the one-sided test based on empirical bootstrap CI is recommended; the fuzzy p-value approach is also desirable, given its theoretical foundation and reasonable performance. For making inference about the uncertainty of $\theta$ estimator, the Wald CI based on the cross-validated estimator is desired for constructing two-sided CI about $\theta$.

\begin{table}[H]
	\centering
	\caption{Na\"{i}ve and cross-validated (CV) estimates of incremental value and corresponding standard deviation (SD) in the parenthesis based on 1000 Monte Carlo simulations under different underlying models and scenarios. $\phi$ indicates the performance of biomarker $X$ alone and $\theta$ indicates the incremental value.}
	\label{Est}
	\begin{tabular}{llllllllllll} \hline
		\multicolumn{12}{c}{Correctly specified model (Logic model)} \\ \hline
		$c_2$ & $c_1$ & $\theta$ & $\phi$ & $\beta_0$ & $\beta_1$ & \multicolumn{3}{c}{Estimate (SD)} & \multicolumn{3}{c}{CV Estimate (SD)} \\
		&  &  &  &  &  & \multicolumn{3}{c}{$n_D=n_{\bar D}$} & \multicolumn{3}{c}{$n_D=n_{\bar D}$} \\
		&  &  &  &  &  & 100 & 250 & 500 & 100 & 250 & 500 \\\hline
		$\infty$ & 0 & 0 & 0.6 & 0.064 & 0.072 & 0.008 & 0.003 & 0.004 & -0.012 & -0.006 & -0.005 \\
		&  &  &  &  &  & (0.010) & (0.004) & (0.004) & (0.025) & (0.010) & (0.011) \\
		$\infty$ & 0 & 0 & 0.7 & 0.028 & 0.144 & 0.002 & 0.001 & 0.000 & -0.008 & -0.003 & -0.002 \\
		&  &  &  &  &  & (0.004) & (0.001) & (0.001) & (0.010) & (0.003) & (0.002) \\
		$\infty$ & 0.84 & 0 & 0.6 & 0.078 & 0.113 & 0.012 & 0.005 & 0.003 & -0.009 & -0.005 & -0.003 \\
		&  &  &  &  &  & (0.005) & (0.006) & (0.003) & (0.028) & (0.012) & (0.007) \\
		$\infty$ & 0.84 & 0 & 0.7 & 0.055 & 0.225 & 0.005 & 0.002 & 0.001 & -0.010 & -0.004 & -0.002 \\
		&  &  &  &  &  & (0.007) & (0.003) & (0.001) & (0.014) & (0.006) & (0.003) \\
		1.15 & 0.84 & 0.05 & 0.6 & 0.061 & 0.129 & 0.054 & 0.051 & 0.051 & 0.040 & 0.047 & 0.048 \\
		&  &  &  &  &  & (0.022) & (0.015) & (0.011) & (0.037) & (0.021) & (0.013) \\
		0.67 & 0.84 & 0.1 & 0.6 & 0.04 & 0.15 & 0.099 & 0.099 & 0.100 & 0.096 & 0.098 & 0.100\\
		&  &  &  &  &  & (0.031) & (0.02) & (0.014) & (0.045) & (0.025) & (0.016) \\
		1.53 & 0.84 & 0.05 & 0.7 & 0.04 & 0.24 & 0.053 & 0.051 & 0.051 & 0.044 & 0.048 & 0.049 \\
		&  &  &  &  &  & (0.019) & (0.012) & (0.009) & (0.027) & (0.014) & (0.010) \\
		1.15 & 0.84 & 0.1 & 0.7 & 0.023 & 0.257 & 0.099 & 0.100 & 0.100 & 0.095 & 0.099 & 0.099 \\
		&  &  &  &  &  & (0.026) & (0.016) & (0.012) & (0.031) & (0.018) & (0.012) \\ \hline
		\multicolumn{12}{c}{Mis-specified model (Logistic model)} \\ \hline
		$\theta$ &  & $\phi$ &  & $\beta_1^*$ & $\beta_2^*$ & \multicolumn{3}{c}{Estimate (SD)} & \multicolumn{3}{c}{CV Estimate (SD)} \\
		&  &  &  &  &  & \multicolumn{3}{c}{$n_D=n_{\bar D}$} & \multicolumn{3}{c}{$n_D=n_{\bar D}$} \\
		&  &  &  &  &  & 100 & 250 & 100 & 250 & 100 & 250 \\ \hline
		0 & & 0.6 & &0.45 & 0 & 0.009 & 0.004 & 0.002 & -0.011 & -0.006 & -0.003 \\
		& & &  &  & &  (0.011) & (0.004) & (0.003) & (0.027) & (0.013) & (0.007) \\
		0 & & 0.7 & &1.1 & 0 & 0.003 & 0.001 & 0.001 & -0.008 & -0.004 & -0.002 \\
		&  & &  &  & & (0.006) & (0.002) & (0.001) & (0.015) & (0.007) & (0.003) \\
		0.05 & & 0.6 & & 0.5 & 0.7 & 0.065 & 0.058 & 0.055 & 0.044 & 0.046 & 0.047 \\
		&  &  & & &  & (0.031) & (0.021) & (0.016) & (0.054) & (0.035) & (0.024) \\
		0.1 & & 0.6 & & 0.5 & 1.05 & 0.108 & 0.106 & 0.104 & 0.093 & 0.099 & 0.100 \\
		&  &  &  &  & & (0.035) & (0.025) & (0.017) & (0.057) & (0.037) & (0.024) \\ \hline
		
	\end{tabular}
\end{table}

\begin{table}[H]
	\centering
	\caption{Type I error rate and power of one-sided test from the empirical bootstrap (EB), percentile bootstrap (PB), Wald using cross-validation (Wald.CV), and fuzzy p-value methods, under different underlying models and scenarios. $\phi$ indicates the performance of biomarker $X$ alone and $\theta$ indicates the incremental value.}
	\label{RejectRate} \small{
	\begin{tabular}{llllllll} \hline
		$c_1$ & $\theta$ & $\phi$ & $n_D=n_{\bar D}$ & EB & PB & Wald.CV & Fuzzy p-value \\ \hline
		\multicolumn{8}{c}{Correctly specified model (Logic model)} \\ \hline
		&  &  &  & \multicolumn{4}{c}{Type I error rate} \\
		0 & 0 & 0.6 & 100 & 0.015 & 0.132 & 0.006 & 0.056 \\
		&  &  & 250 & 0.007 & 0.135 & 0.003 & 0.034 \\
		&  &  & 500 & 0.023 & 0.344 & 0.008 & 0.055 \\
		&  & 0.7 & 100 & 0.001 & 0.009 & 0.001 & 0.038 \\
		&  &  & 250 & 0.005 & 0.006 & 0.001 & 0.042 \\
		&  &  & 500 & 0.003 & 0.011 & 0.001 & 0.047 \\
		0.84 &  & 0.6 & 100 & 0.033 & 0.229 & 0.011 & 0.073 \\
		&  &  & 250 & 0.015 & 0.256 & 0.002 & 0.042 \\
		&  &  & 500 & 0.018 & 0.277 & 0.004 & 0.036 \\
		&  & 0.7 & 100 & 0.013 & 0.077 & 0.004 & 0.043 \\
		&  &  & 250 & 0.012 & 0.087 & 0.000 & 0.041 \\
		&  &  & 500 & 0.006 & 0.078 & 0.001 & 0.042 \\ 
		&  &  &  & \multicolumn{4}{c}{Power} \\
		& 0.05 & 0.6 & 100 & 0.593 & 0.905 & 0.255 & 0.426 \\
		&  &  & 250 & 0.945 & 0.996 & 0.710 & 0.690 \\
		&  &  & 500 & 0.998 & 1.000 & 0.969 & 0.889 \\
		& 0.1 &  & 100 & 0.916 & 0.988 & 0.713 & 0.836 \\
		&  &  & 250 & 1.000 & 1.000 & 0.991 & 0.984 \\
		&  &  & 500 & 1.000 & 1.000 & 1.000 & 1.000 \\
		& 0.05 & 0.7 & 100 & 0.801 & 0.942 & 0.499 & 0.462 \\
		&  &  & 250 & 0.990 & 0.999 & 0.949 & 0.772 \\
		&  &  & 500 & 1.000 & 1.000 & 0.997 & 0.945 \\
		& 0.1 &  & 100 & 0.991 & 0.997 & 0.908 & 0.890 \\
		&  &  & 250 & 1.000 & 1.000 & 1.000 & 0.995 \\
		&  &  & 500 & 1.000 & 1.000 & 1.000 & 1.000 \\ \hline
		\multicolumn{8}{c}{Mis-specified model (Logistic model)} \\ \hline
		&  &  &  & \multicolumn{4}{c}{Type I error rate} \\
		NA & 0 & 0.6 & 100 & 0.019 & 0.144 & 0.009 & 0.069 \\
		&  &  & 250 & 0.021 & 0.144 & 0.011 & 0.057 \\
		&  &  & 500 & 0.008 & 0.167 & 0.008 & 0.041 \\
		&  & 0.7 & 100 & 0.008 & 0.026 & 0.005 & 0.048 \\
		&  &  & 250 & 0.007 & 0.025 & 0.006 & 0.041 \\
		&  &  & 500 & 0.013 & 0.031 & 0.004 & 0.044 \\
		&  &  &  & \multicolumn{4}{c}{Power} \\
		& 0.05 & 0.6 & 100 & 0.512 & 0.935 & 0.212 & 0.466 \\
		&  &  & 250 & 0.755 & 0.999 & 0.382 & 0.647 \\
		&  &  & 500 & 0.924 & 1.000 & 0.627 & 0.812 \\
		& 0.1 &  & 100 & 0.870 & 0.996 & 0.519 & 0.796 \\
		&  &  & 250 & 0.990 & 1.000 & 0.871 & 0.968 \\
		&  &  & 500 & 1.000 & 1.000 & 0.991 & 0.996 \\ \hline
	\end{tabular} }
\end{table}

\begin{table}[H]
	\centering
	\caption{Coverage of 95\% two-sided confidence interval (CI) and corresponding length in the parenthesis using Wald with cross-validation method, under different underlying models and scenarios. $\phi$ indicates the performance of biomarker $X$ alone and $\theta$ indicates the incremental value.}
	\label{Coverage} 
	\begin{tabular}{llllll} \hline
		$c_1$ & $\theta$ & $\phi$ & \multicolumn{3}{c}{Coverage (Length) of two-sided 95\% CI} \\
		&  &  & \multicolumn{3}{c}{$n_D=n_{\bar D}$} \\
		&  &  & 100 & 250 & 500 \\ \hline
		\multicolumn{6}{c}{Correctly specified model (Logic model)} \\ \hline
		0 & 0 & 0.6 & 98.3\% (0.117) & 99.9\% (0.053) & 97.8\% (0.051) \\
		&  & 0.7 & 98.0\% (0.056) & 96.8\% (0.019) & 95.9\% (0.008) \\
		0.842 &  & 0.6 & 98.5\% (0.123) & 98.8\% (0.060) & 98.2\% (0.034) \\
		&  & 0.7 & 97.3\% (0.017) & 97.9\% (0.030) & 97.8\% (0.015) \\
		& 0.05 & 0.6 & 94.3\% (0.154) & 96.0\% (0.089) & 96.5\% (0.056) \\
		& 0.1 & 0.6 & 95.4\% (0.175) & 95.7\% (0.100) & 96.9\% (0.065) \\
		& 0.05 & 0.7 & 93.8\% (0.108) & 96.4\% (0.060) & 95.5\% (0.039) \\
		& 0.1 & 0.7 & 95.3\% (0.128) & 95.6\% (0.074) & 95.7\% (0.049) \\ \hline
		\multicolumn{6}{c}{Mis-specified model (Logistic model)} \\ \hline
		NA & 0  & 0.6 & 98.4\% (0.123) & 97.9\% (0.063) & 98.0\% (0.035) \\
		& 0 & 0.7 & 97.4\% (0.072) & 97.7\% (0.032) & 98.6\% (0.017) \\
		& 0.05  & 0.6 & 93.7\% (0.202) & 92.9\% (0.132) & 94.2\% (0.094) \\
		& 0.1 & 0.6 & 92.9\% (0.215) & 93.2\% (0.137) & 95.1\% (0.097) \\ \hline
	\end{tabular}
\end{table}

\section{Pancreatic Cancer Study}
In this section, we apply the proposed methods to a real data example from a pancreatic cancer study for identifying biomarkers for early detection of pancreatic cancer. In this study, plasma samples were collected from $n_{D}=109$ patients with pancreatic cancer and $n_{\bar{D}}=91$ healthy individuals for biomarker measurement \citep{Tang2015}. The sialyl-Lewis A (sLeA) glycan, on which the CA19-9 assay is based, is currently the only established biomarker for pancreatic cancer detection. However, its performance for early detection of pancreatic cancer is not satisfactory given that it is not evaluated in about 25\% of pancreatic cancers. \cite{Tang2015} found that sialyl-Lewis X (sLeX), a structural isomer of sLeA, was elevated in the plasma of 14-19\% of patients with low sLeA and thus a biomarker panel combining sLeA and sLeX can potentially be useful in the clinical detection of pancreatic cancer. In this study, the estimated optimal average sensitivity and specificity based on sLeA alone is 0.683 (0.636 after cross-validation). Here, we estimate the incremental value of sLeX when combined with sLeA using an OR rule and test the hypothesis that a strategy combining the two biomarkers performs better than using the sLeA biomarker alone. The estimated na\"{i}ve and cross-validated incremental value of sLeX is 0.079 and 0.062, respectively. We apply the EB, PB, and Wald.CV methods to construct one-sided confidence intervals for the incremental value of sLeX. We also apply the fuzzy p-value approach to the data and compute the average rejection rate over 1000 random draws. In addition, the two-sided CI based on Wald.CV is constructed. 

When using the original data set, the one-sided bootstrap CIs based on EB and PB provide strong evidence against the null hypothesis while the Wald.CV method fails to do so, which is not surprising given that EB and PB have been shown to be more powerful compared to Wald.CV. The fuzzy p-value approach rejects the null hypothesis with probability 0.67. The two-sided CI derived using Wald.CV is $\left[-0.030,0.153\right]$. 

Moreover, to investigate the impact of increased sample size, we generate a larger dataset by randomly drawing 200 cases and 200 controls with replacement from the original data and apply our proposed methods. With the larger sample sizes, CIs based on EB, PB, and Wald.CV all have lower limits above zero, providing strong evidence that adding sLeX yields significantly better performance compared to using sLeA alone. The fuzzy p-value approach rejects the null hypothesis with probability 1. The Wald.CV two-sided CI turns out to be $\left[0.066, 0.164 \right]$.

\section{Concluding Remarks}
In this paper, we considered an inference problem about the incremental value of a new biomarker when combined with an established biomarker using an OR rule, motivated by the example in early detection of pancreatic cancer, where the standard biomarker CA19-9 is only elevated in a subclass of cancer cases. Thus, identifying a new biomarker that is present in the other subclasses to combine with CA19-9 is of primary interest. We considered a nonparametric estimator of incremental value of the new biomarker, based on an estimator of the OR rule that maximizes the weighted average of sensitivity and specificity. We proposed different procedures based on bootstrap, cross-validation and a novel fuzzy p-value approaches, to test and make inference about a new biomarker's incremental value. Through extensive numerical studies, we found that the hypothesis test based on one-sided empirical bootstrap CI has satisfactory performance in terms of well-controlled type I error rate and decent power for declaring the usefulness of the new marker, while the popular percentile bootstrap CI should be avoided due to its inflated type I error rate. When it is of interest to provide uncertainty about the estimated incremental value, we found that two-sided Wald-type CI based on cross-validated estimates of incremental value performs very well with the coverage closed to the nominal level. The novel fuzzy p-value method we proposed for testing the incremental value also has satisfactory performance. Moreover, the fuzzy p-value method can be particularly appealing as a testing procedure given its theoretical foundation and its potential to be extended to other biomarker testing problems when non-regularity is an issue. Importantly, our findings are based not only on settings where the true risk model conditional on biomarkers follows a logic model with an OR combination, but also on settings where the logic risk model does not hold but the OR rule is used as a practical way to combine biomarkers for simplicity and interpretability. Our findings provide valuable guidance on selecting appropriate methods for testing and making inference about the incremental value of a new biomarker.

\section*{Acknowledgements}
This work was supported by the U.S. National Institutes of Health under award R01
GM106177-01. 

\section*{Appendix}

\begin{table}[]
	\centering
	\caption{Na\"{i}ve and cross-validated (CV) estimates of incremental value with corresponding standard deviation (SD) in the parenthesis, power of one-sided test from the empirical bootstrap (EB), percentile bootstrap (PB), Wald using cross-validation (Wald.CV), and fuzzy p-value methods, and coverage of 95\% two-sided confidence interval (CI) with corresponding length in the parenthesis using Wald.CV method, under the local alternative $\theta=0.05\sqrt{\frac{100}{n_{D}}}$, or $0.1\sqrt{\frac{100}{n_{D}}}$, based on 1000 Monte Carlo simulations. $\phi$ indicates the performance of biomarker $X$ alone and $\theta$ indicates the incremental value.}
	\begin{tabular}{llllllllll} \hline
$\phi$ & $\theta$ & $n_D$ & Estimate  & CV Estimate   & \multicolumn{4}{c}{Power} & Coverage  \\
& & & (SD) & (SD)  & &&& & (Length) \\
&  &  &  &  & EB & PB & Wald.CV & Fuzzy & of two-sided \\
&  &  &  &  &  &  &  & p-value & 95\% CI \\ \hline
0.6 & 0.05 & 100 & 0.054 & 0.040 & 0.593 & 0.905 & 0.255 & 0.426 & 94.30\% \\
&  &  & (0.022) & (0.037) &  &  &  &  & (0.154) \\
& 0.03 & 250 & 0.034 & 0.027 & 0.773 & 0.985 & 0.366 & 0.434 & 95.00\% \\
&  &  & (0.012) & (0.019) &  &  &  &  & (0.080) \\
& 0.02 & 500 & 0.024 & 0.019 & 0.882 & 0.993 & 0.520 & 0.421 & 96.20\% \\
&  &  & (0.007) & (0.011) &  &  &  &  & (0.047) \\
& 0.1 & 100 & 0.099 & 0.096 & 0.916 & 0.988 & 0.713 & 0.836 & 95.40\% \\
&  &  & (0.031) & (0.045) &  &  &  &  & (0.175) \\
& 0.06 & 250 & 0.064 & 0.062 & 0.979 & 0.999 & 0.862 & 0.847 & 95.40\% \\
&  &  & (0.017) & (0.023) &  &  &  &  & (0.093) \\
& 0.04 & 500 & 0.046 & 0.043 & 0.994 & 1.000 & 0.933 & 0.837 & 96.30\% \\
&  &  & (0.010) & (0.013) &  &  &  &  & (0.055) \\
0.7 & 0.05 & 100 & 0.053 & 0.044 & 0.801 & 0.942 & 0.499 & 0.462 & 93.80\% \\
&  &  & (0.019) & (0.027) &  &  &  &  & (0.108) \\
& 0.03 & 250 & 0.033 & 0.028 & 0.914 & 0.992 & 0.705 & 0.445 & 94.60\% \\
&  &  & (0.010) & (0.013) &  &  &  &  & (0.052) \\
& 0.02 & 500 & 0.023 & 0.020 & 0.976 & 1.000 & 0.881 & 0.461 & 95.20\% \\
&  &  & (0.006) & (0.007) &  &  &  &  & (0.030) \\
& 0.1 & 100 & 0.099 & 0.095 & 0.991 & 0.997 & 0.908 & 0.890 & 95.30\% \\
&  &  & (0.026) & (0.031) &  &  &  &  & (0.128) \\
& 0.06 & 250 & 0.064 & 0.061 & 0.999 & 1.000 & 0.990 & 0.890 & 95.10\% \\
&  &  & (0.014) & (0.016) &  &  &  &  & (0.065) \\
& 0.04 & 500 & 0.045 & 0.043 & 1.000 & 1.000 & 0.999 & 0.898 & 95.10\% \\
&  &  & (0.009) & (0.009) &  &  &  &  & (0.038) \\ \hline
	\end{tabular}
\end{table}
\end{document}